\documentclass[twocolumn,english,prl,showpacs,preprintnumbers,superscriptaddress]{revtex4-1}
\usepackage[utf8]{inputenc}
\usepackage{amsmath}
\usepackage{amssymb}
\usepackage{graphicx}
\usepackage{textcomp}
\usepackage{bm}
\usepackage[right]{eurosym}
\usepackage{float}
\usepackage[english]{babel}
\usepackage{blindtext}
\usepackage{babel}
\usepackage{lineno}
\usepackage{booktabs}
\usepackage{longtable}

\begin{document}
	\author{R. Michiels}
	\email{rupert.michiels@physik.uni-freiburg.de}	
	\affiliation{Institute of Physics, University of Freiburg, 79104 Freiburg, Germany}
	\author{A. C. LaForge}
	\affiliation{Institute of Physics, University of Freiburg, 79104 Freiburg, Germany}
	\affiliation{Department of Physics, University of Connecticut, Storrs, Connecticut, 06269, USA}
	\author{M. Bohlen}
	\affiliation{Institute of Physics, University of Freiburg, 79104 Freiburg, Germany}
	\author{C. Callegari}
	\affiliation{Elettra-Sincrotrone Trieste, 34149 Basovizza, Trieste, Italy}
	\author{A. Clark}
	\affiliation{Laboratory of Molecular Nanodynamics, Ecole Polytechnique F{\'e}d{\'e}rale de Lausanne, 1015 Lausanne, Switzerland}
	\author{A. von Conta}
	\affiliation{Laboratorium f{\"u}r Physikalische Chemie, ETH Z\"urich, 8093 Z\"urich, Switzerland}
	\author{M. Coreno}
	\affiliation{ISM-CNR, Istituto di Struttura della Materia, LD2 Unit, 34149 Trieste, Italy}
	\author{M. Di Fraia}
	\affiliation{Elettra-Sincrotrone Trieste, 34149 Basovizza, Trieste, Italy}
	\author{M. Drabbels}
	\affiliation{Laboratory of Molecular Nanodynamics, Ecole Polytechnique F{\'e}d{\'e}rale de Lausanne, 1015 Lausanne, Switzerland}
	\author{P. Finetti}
	\affiliation{Elettra-Sincrotrone Trieste, 34149 Basovizza, Trieste, Italy}
	\author{M. Huppert}
	\affiliation{Laboratorium f{\"u}r Physikalische Chemie, ETH Z\"urich, 8093 Z\"urich, Switzerland}
	\author{V. Oliver}
	\affiliation{Laboratory of Molecular Nanodynamics, Ecole Polytechnique F{\'e}d{\'e}rale de Lausanne, 1015 Lausanne, Switzerland}
	\author{O. Plekan}
	\affiliation{Elettra-Sincrotrone Trieste, 34149 Basovizza, Trieste, Italy}
	\author{K. C. Prince}
	\affiliation{Elettra-Sincrotrone Trieste, 34149 Basovizza, Trieste, Italy}
	\author{S. Stranges}
	\affiliation{Department of Chemistry and Drug Technologies, University Sapienza, 00185 Rome, Italy, and Tasc IOM-CNR, Basovizza, Trieste, Italy}
	\author{V. Svoboda}
	\affiliation{Laboratorium f{\"u}r Physikalische Chemie, ETH Z\"urich,	 8093 Z\"urich, Switzerland}
	\author{H. J. W\"orner}
	\affiliation{Laboratorium f{\"u}r Physikalische Chemie, ETH Z\"urich,	 8093 Z\"urich, Switzerland}
	\author{F. Stienkemeier}
	\affiliation{Institute of Physics, University of Freiburg, 79104 Freiburg, Germany}

	\title[An \textsf{achemso} demo]
	{Time-resolved formation of excited atomic and molecular states in XUV-induced nanoplasmas in ammonia clusters}
	\begin{abstract}
		High intensity XUV radiation from a free-electron (FEL) was used to create a nanoplasma inside ammonia clusters with the intent of studying the resulting electron-ion interactions and their interplay with plasma evolution. In a plasma-like state, electrons with kinetic energy lower than the local collective Coulomb potential of the positive ionic core are trapped in the cluster and take part in secondary processes (e.g. electron-impact excitation/ionization and electron-ion recombination) which lead to subsequent excited and neutral molecular fragmentation. Using a time-delayed UV laser, the dynamics of the excited atomic and molecular states are probed from -0.1\,ps to 18\,ps. We identify three different phases of molecular fragmentation that are clearly distinguished by the effect of the probe laser on the ionic and electronic yield. We propose a simple model to rationalize our data and further identify two separate channels leading to the formation of excited hydrogen.
	\end{abstract}
	\date{\today}
	\maketitle
	
\section{Introduction}
Through the absorption of intense radiation, plasma-like states can be formed in condensed systems within a few femtoseconds. In particular, the study of laser-induced nanoplasmas in rare gas clusters has attracted considerable interest~\cite{fennel2010laser,saalmann2006mechanisms} over a wide range of wavelengths, from strong-field ionization in the infrared to multiphoton ionization in the X-ray. Specifically for short wavelength radiation, the mechanism of nanoplasma formation is as follows: first, electron emission is due to direct single photon absorption until a positive ionic core is built up within the cluster. Afterwards, electrons are ionized from individual atoms in the cluster, but are trapped by the surrounding electrostatic cluster potential leading to frustration in the emission process and a plasma-like state being formed.~\cite{wabnitz2002multiple,bostedt2008multistep,arbeiter2011rare,arbeiter2010ionization,iwayama2013frustration} At later times, the ions in the nanoplasma move apart under the combined action of  hydrodynamic forces and of Coulomb repulsion. Previous research on nanoplasmas, focusing primarily on rare gas clusters, has shown that frustration of electron emission can indeed be achieved with intense XUV pulses, and that inelastic collisions and recombination processes play an important role in the nanoplasma evolution.~\cite{jungreuthmayer2005intense,muller2015ionization,schutte2014tracing,oelze2017correlated,schutte2017tracing,pohl2008rydberg,iwayama2015transient,schutte2015real}\\ 

For molecular clusters, as compared to their atomic counterparts, the presence of both intra- and intermolecular bonds complicate the nanoplasma dynamics and the underlying fragmentation processes. Specifically, the ejection of lightweight hydrogen in molecular clusters offers an efficient pathway for both cooling and charge dissipation.~\cite{di2013proton,kaplan2003shock,last2001nuclear} The study of highly ionized molecular clusters has previously been investigated using intense radiation from a tabletop laser~\cite{snyder1996femtosecond,card1997covariance,niu2005covariance,wang2008cluster} and a FEL,~\cite{iwan2012explosion} but, so far, those results fail to resolve the temporal evolution of the underlying mechanisms and, therefore, little is known about the ultrafast nanoplasma dynamics. We report on time-resolved measurements on ammonia clusters using intense, femtosecond XUV FEL pulses to create a highly-ionized, plasma-like state. Using a 266\,nm UV laser, we probed the evolution of the nanoplasma with a specific focus on the  atomic and molecular excited states. In the ion and electron yields, we observe the formation of excited hydrogen as a major contribution (Fig.\,\ref{fig:2}). From the time-resolved data we can identify two dominant pathways for the creation of excited hydrogen. These can be created either via direct recombination of ions and electrons, or by dissociation of highly excited molecules.\\ 
\begin{figure}[ht]
	\begin{center}
		\includegraphics[width=\columnwidth]{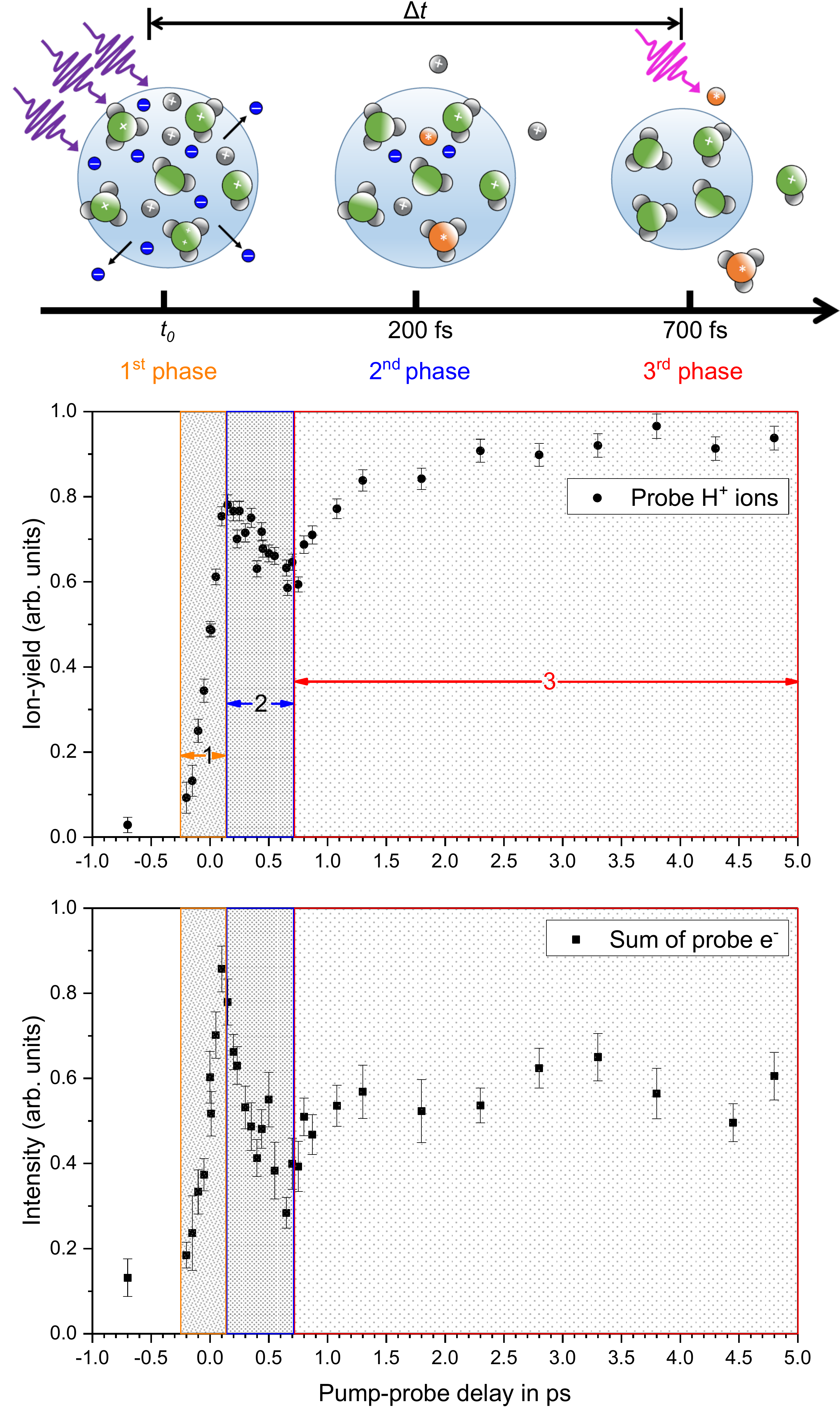}
		\caption{Top: Schematic of the cluster explosion illustrating the three phases. Middle: Probe yield of H$^+$ ions with varying pump-probe delay. The shaded areas indicate the three phases of the cluster explosion as discussed in the text: 1) Direct pulse overlap phase. 2) Recombination and proton-radiation phase. 3) Molecular dissociation phase. Bottom: Absolute sum of the probe photoelectron yield.}
		\label{fig:1}
	\end{center}
\end{figure}
\section{Experimental Setup and Methods}
The experiment was performed at the Low Density Matter (LDM) endstation~\cite{lyamayev2013modular} at the seeded FEL FERMI in Trieste, Italy.~\cite{allaria2012highly,allaria2012tunability} The photon energy was set to $h\nu$ = 24.0\,eV and the pulse length was approximately 70\,fs FWHM. The FEL pulse energy at the target was approximately 70\,$\mu$J, calculated from the value measured upstream on a shot-by-shot basis by gas ionization and the nominal reflectivity of the optical elements in the beam transport system. The FEL photon energy of 24.0\,eV was chosen in order to maximize the product of photon flux and ionization cross section within the technical limitations posed by the FEL. The FEL spot size in the interaction region was 20\,$\mu$m FWHM. The pump-probe scheme was realized using a UV pulse produced from a frequency-tripled Ti:Sapphire laser ($h\nu$ = 4.75\,eV) with a pulse energy of about  45\,$\mu$J and a spot size of 80\,$\mu$m FWHM. The cross-correlation between the two pulses, measured by resonant two photon ionization of helium, was about 200\,fs FWHM. A supersonic jet of ammonia clusters was produced by expansion of pressurized ammonia through a custom, pulsed nozzle. By varying the backing pressure of the ammonia, we controlled the mean cluster size in the range of 10$^1$ -- 3$\times$10$^3$ ammonia molecules. For this experiment, a mean cluster size of 2000 molecules was chosen giving the best signal. The mean cluster sizes where determined using a modified scaling law of the Hagena type.~\cite{bobbert2002fragmentation} The cluster beam was perpendicularly crossed by the FEL/UV beam at the focus of a velocity-map-imaging (VMI) spectrometer, and of an ion time-of-flight mass spectrometer.~\cite{lyamayev2013modular} The electron kinetic energy and angular distributions were reconstructed from velocity-map images using the Maximum Entropy Legendre Reconstruction method.~\cite{dick2014inverting} To emphasize the prevalence of cluster processes as opposed to processes involving isolated molecules we compare our results with the interaction of XUV light with molecular gas phase ammonia. These interactions are well studied experimentally and theoretically and the partial ionization cross sections leading to different ionic fragments are known from experiment~\cite{samson1987absorption} and calculations.~\cite{li2010ab,biesner1989state, bach2003competition}\\

\section{Results and Discussion}

\subsection{Time-resolved probe ion yields}

Irradiating a gas jet of molecular ammonia employing the above-stated FEL pulse parameters results in a sample ionization rate that is near saturation ($>$90\%). Considering similar absorption rates in ammonia clusters results in the system having multiple ions in a single cluster, which leads to the formation of a nanoplasma. To understand the dynamics of a molecular nanoplasma and motivate our three phase interpretation, we initially examine the time dependence of the probe ion yields and specifically focus on H$^+$ ions since it is known they play a strong role in the evolutionary behavior of molecular clusters.~\cite{di2013proton} The probe yield of H$^+$ ions as a function of pump-probe delay is shown in Fig.\,\ref{fig:1}. The proton yield for $t<t_0$ is close to zero, since the 266\,nm pulse, due to low intensity, cannot directly create a significant fraction of H$^+$ ions. In the pulse overlap, around $t = 0$, the probe yield drastically increases, and the observed half-width resembles the cross correlation of the two laser pulses, shown in the supplementary information. This is the first phase where the cluster is being multiply ionized leading to the formation of a nanoplasma. As sequential ionization proceeds, quasi-free electrons are trapped by the positive cluster potential leading to a frustration in ionization, a defining property of XUV-induced nanoplasmas.~\cite{bostedt2008multistep,arbeiter2011rare,arbeiter2010ionization,iwayama2013frustration} To characterize this behavior, we introduce the frustration parameter, which can be estimated as $\alpha = N_{tot}/q_{full}$, the ratio of total ionization events to the number of primary ionizations prior to full frustration $q_{full} = (\hbar\omega-I_p)r_sN^{1/3}/(1.44$\,eV\,\AA$)$ ($r_s$ the Wigner-Seitz radius) for a spherical homogeneously charged cluster.~\cite{arbeiter2011rare} The frustration parameter for this experiment is $\alpha >15$. Since the frustration parameter is much larger than one, electrons are trapped inside the cluster potential and hence recombination with the ions is strongly enhanced. After ionization frustration sets in, the second phase begins where the H$^+$ probe yield decreases linearly reaching a minimum value at roughly 700\,fs pump-probe delay. We attribute the decrease to the quick decay of excited molecular states which are populated by electron-ion recombination and molecular formation within the cluster. A similar behavior was observed for the fragmentation of small excited ammonia clusters and attributed to the lifetime of the excited \~{A} state of ammonia.~\cite{freudenberg1996ultrafast} Furthermore, charge equalization takes place during the second phase, as protons leave the cluster and carry away the excess charge. Electrons that have not recombined with ions will successively leave the cluster as low kinetic energy electrons once the local potential barrier is lower than the kinetic energy of the electrons. Eventually, there is little excess charge in the cluster and the third phase follows. In phase three the H$^+$ probe yield shown in Fig.\,\ref{fig:1} increases, converging asymptotically to a higher final value than the intermediate maximum. The third phase is dominated by relaxation and dissociation processes of excited free molecules, atoms and cluster fragments. Complimentary to the H$^+$ ion yields, the sum of the probe electron yield, in the bottom panel of Fig.\,\ref{fig:1}, shows identical behavior further prompting our three phase model interpretation. Additionally, similar behavior is observed in many other ion yields, which are shown in Fig.\,\ref{fig:4} and in the supplementary information~\textsuperscript{\dag}.\\ 

\subsection{Time-resolved probe photoelectrons}

\begin{figure}[t]
	\centering
	\includegraphics[width=\columnwidth]{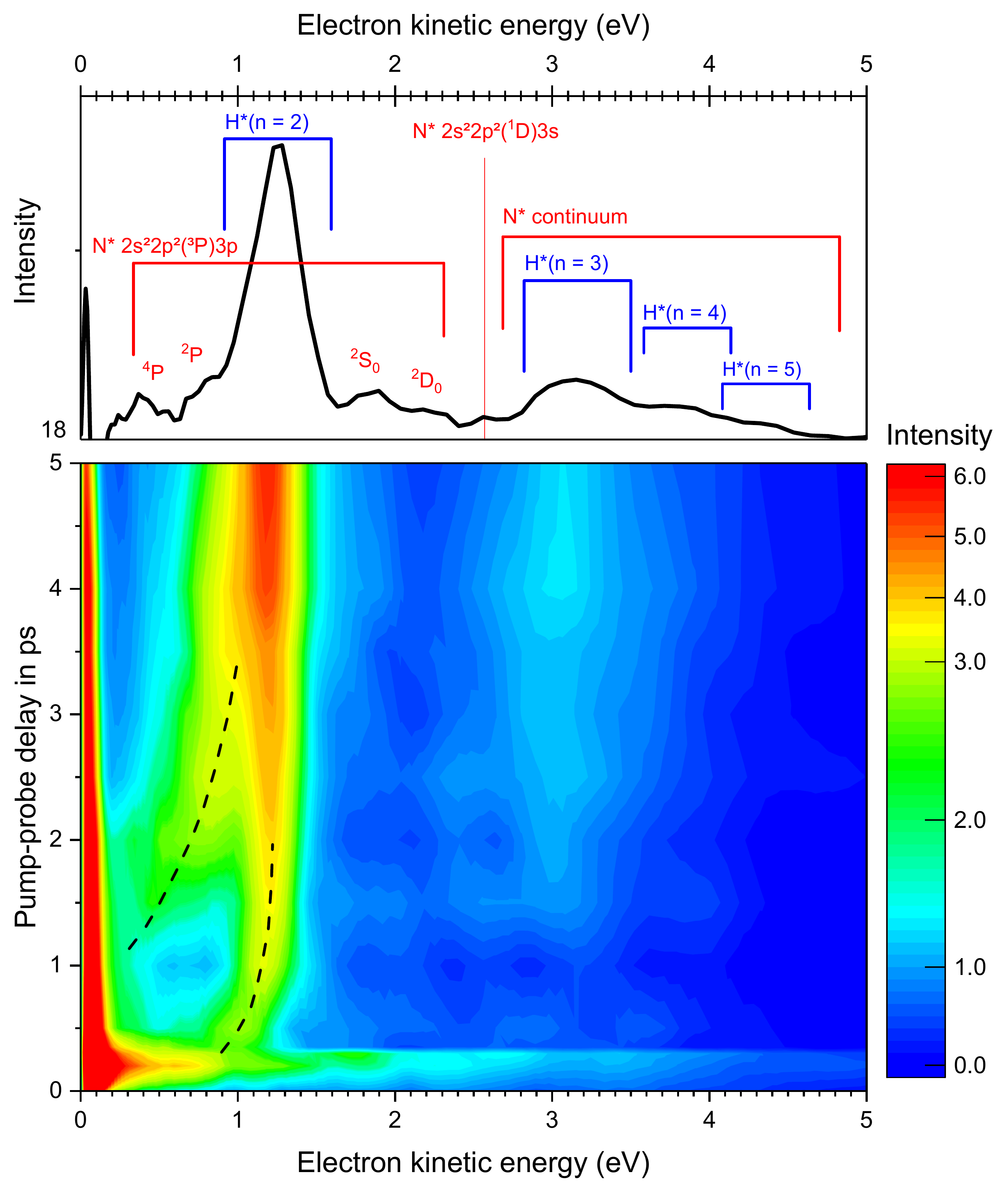}
	\caption{Top: Probe photoelectron intensities for ammonia clusters are shown versus electron kinetic energy at an XUV-UV pump-probe delay of 18\,ps and 24.0\,eV FEL photon energy. The contributions from excited hydrogen states and their expected binding energies are marked in blue and the excited nitrogen states are marked in red. Bottom: Color plot of the probe photoelectron intensities with pump-probe delay on the y-axis and electron kinetic energy on the x-axis. Dashed black lines are used to highlight the first and second formation channel of H$^*$.}
	\label{fig:2}
\end{figure}

Next, we focus on the UV-ionized probe electron spectrum (UV-PES) at a time delay where the ultrafast electronic and nuclear dynamics have ended. The UV-PES, shown in the top panel of Fig.\,\ref{fig:2} contains only the probe photoelectrons which were extracted by subtracting an XUV only PES from a XUV+UV PES at a pump-probe delay of 18\,ps. The prominent feature at $1.30\pm0.05$\,eV has a narrow width of $0.30\pm0.05$\,eV FWHM, primarily due to the experimental resolution. From the vertical binding energy, and comparison with our previous experiments, we know that these electrons come from excited hydrogen in the n\,=\,2 state.~\cite{laforge2019real} We also observe higher lying excited states of hydrogen, with kinetic energy\,$>$\,3\,eV, but with less intensity. The recombination coefficients of electrons and protons leading to the formation of excited hydrogen with different principle quantum number n can be calculated from quantum mechanics~\cite{bates2012atomic,burgess1965tables} and depend on the relative collision energy ($E_{coll}$). Taking the ionization cross section for the 266\,nm probe into account, we expect the relative observed photoelectron intensities for excited hydrogen with principal quantum number n\,=\,2/n\,=\,3/n\,=\,4 to be 1/0.24/0.08 when $E_{Coll}= 0$ and 1/0.11/0.02 when $E_{Coll}= 15$\,eV. The relative intensities in our experiment are 1/0.18$\pm$0.06/0.09$\pm$0.07, in between the two limits, as is expected since electron-proton recombination in our experiment is occurring with electron kinetic energies between 0 and 15\,eV. Furthermore, as we will discuss, the excited hydrogen in our experiment are not exclusively created by recombination of electrons and protons. Analysis of the remaining photoelectron features in the top panel of Fig.\,\ref{fig:2} shows a large overlap with the excited states of nitrogen. This observation is further supported by the probe ion yield\textsuperscript{\dag}, which is dominated by protons (86\%) and nitrogen ions (7\%). This suggests that a major part of the remaining photoelectrons comes from the excited states of atomic nitrogen. A contribution from solvated electrons as observed in~\cite{laforge2019real} would be observed at 3\,eV electron kinetic energy, which strongly overlaps with photoelectrons from excited hydrogen and nitrogen. For this reason, a contribution from solvated electrons can neither be excluded nor included.\\

\begin{figure*}
	\begin{center}
		\includegraphics[width=2\columnwidth]{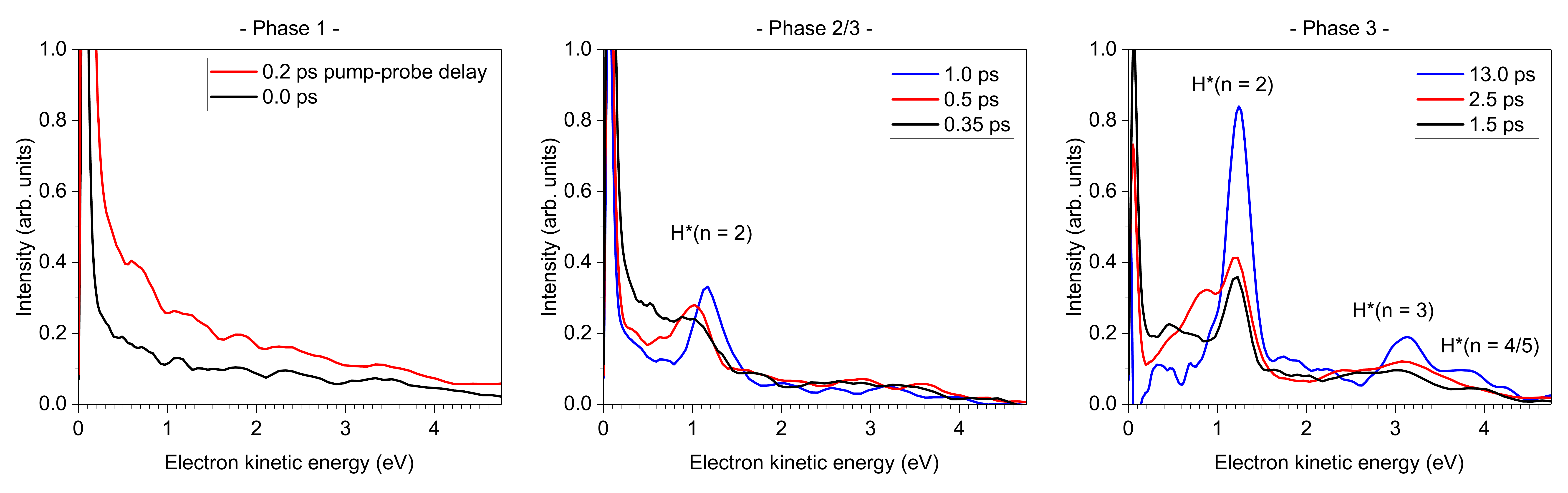}
		\caption{Selected cuts from the probe photoelectron yield. Left Panel: Probe photoelectrons for zero and 200\,fs pump-probe delay. Middle Panel: Pump-probe delay from 350\,fs to 1\,ps. Right Panel: Pump-probe delay $>$\,1\,ps.}
		\label{fig:3}
	\end{center}
\end{figure*}
Complementary to the ion yields, shown in Fig.\,\ref{fig:1}, the time-resolved UV-PES gives information about the evolution of the respective electronic binding potentials. Horizontal projections of the lower panel of Fig.\,\ref{fig:2} for different pump-probe delay are shown in separate panels in Fig.\,\ref{fig:3}. Initially, within the first few hundreds of femtoseconds, the UV-PES is dominated by a large peak of low kinetic energy electrons with a broad distribution extending to higher energies; no additional peaks are observed since the cluster potential screens the binding energies of the ionized states. The observed features reach maximum intensity at 150\,fs pump-probe delay, coinciding with the end of phase one. Around a pump-probe delay of 500\,fs, shown as red curve in the middle panel of Fig.\,\ref{fig:3}, we observe the first well-defined peak from the n\,=\,2 state of excited hydrogen. The peak is slightly offset from its nominal value by 300\,meV. In the following 500\,fs the peak converges to the kinetic energy precisely matching the vertical binding energy of H$^*$(n\,=\,2) (blue curve). In addition to initially being offset, the peak is also broadened; both of these features indicate the influence of the local cluster potential on the atomic hydrogen binding energy. As the peak emerges immediately with the decaying nanoplasma, it seems reasonable to assign this first channel to a direct recombination of protons and electrons in the nanoplasma. In the further evolution of the UV-PES in the bottom panel of Fig.\,\ref{fig:2}, we can see a new photoelectron line appear at roughly 1.2\,ps pump-probe delay. The kinetic energy of these electrons is at first $<$ 1\,eV and the peak is clearly separated from the fully developed H$^*$(n\,=\,2) photoelectron peak as shown in the black and red curve in the right panel of Fig.\,\ref{fig:3}. In the following picoseconds, the new peak gains in intensity and converges to the expected kinetic energy for excited hydrogen photoelectrons. From the clear differences in the timescale, we conclude that two separate processes forming excited hydrogen are observed. \\

\subsection{Formation of excited hydrogen by molecular dissociation}
In this section, we discuss the second contribution of the observed excited hydrogen using rigorous analysis of the photoelectron and ion time-of-flight spectra. A first step will be to identify possible precursor molecules or cluster fragments from which the H$^*$ dissociates. The time-resolved probe ion yields of NH$_4^+$, the dimer, and the larger cluster fragments are shown in Fig.\,S4 in the supplementary information\textsuperscript{\dag}. Despite the considerable temporal evolution in the first and second phase of the nanoplasma, the probe ion yield of all larger fragments stays constant in the third phase. From this observation we conclude that the precursor molecule is not a cluster fragment (NH$_3$)$_n$H$_m$ with $n>1$ and $m$ any integer. The next important observation is that the convergence of the photoelectron kinetic energy, as illustrated by dashed black lines in Fig.\,\ref{fig:2}, evolves from lower kinetic energy towards higher kinetic energy. We estimate that the first clear observation of photoelectrons from the second channel is at 1.4\,ps pump-probe delay, with a mean electron kinetic energy of 400\,meV. This is shifted by 900\,meV compared to the final convergence value of 1.3\,eV. Most neutrally dissociating potential energy surfaces of molecules evolve in the direction of higher electron binding energy.~\cite{davies1999femtosecond,meier2002time,tsubouchi2005ultrafast} The rising kinetic energy observed in the experimental spectrum indicates that the ionization energy is dropping during the dissociation. Since the observed change is too large to be explained by dissociation of any neutral ammonia fragment, we deduce that the precursor is a positively charged molecule. In conclusion, the precursor is a molecule which dissociates into unknown fragments plus H$^*$, with at least one of the unknown fragments being ionic.\\

\begin{figure}
	\begin{center}
		\includegraphics[width=\columnwidth]{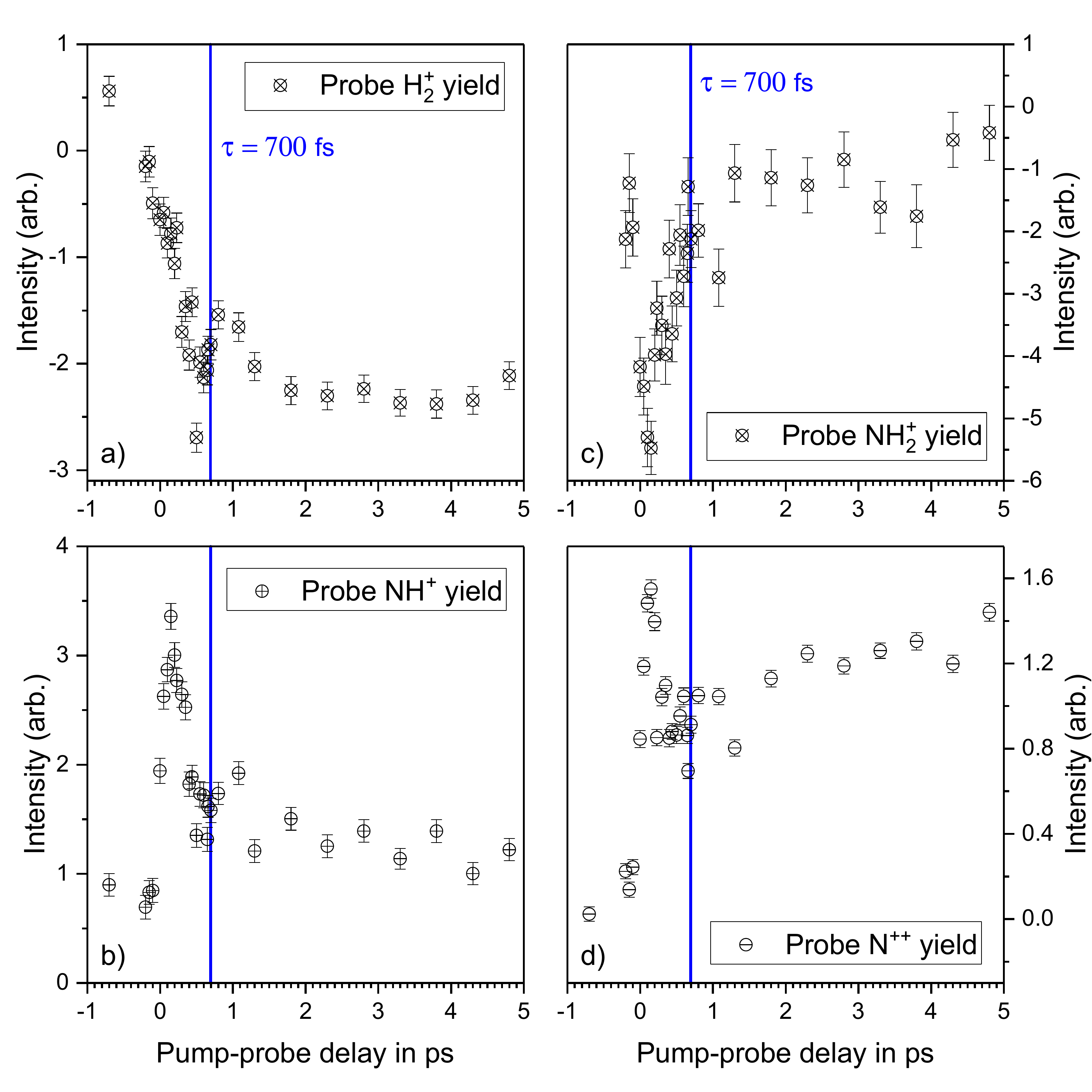}
		\caption{Probe ion yields versus pump-probe delay for H$_2^+$ (a), NH$^{+}$ (b), NH$_2^+$ (c), and N$^{++}$ (d). The blue vertical line at 700\,fs pump-probe delay marks the transition between the second and third phase of the nanoplasma expansion.}
		\label{fig:4}
	\end{center}
\end{figure}

\begin{figure}
	\begin{center}
		\includegraphics[width=\columnwidth]{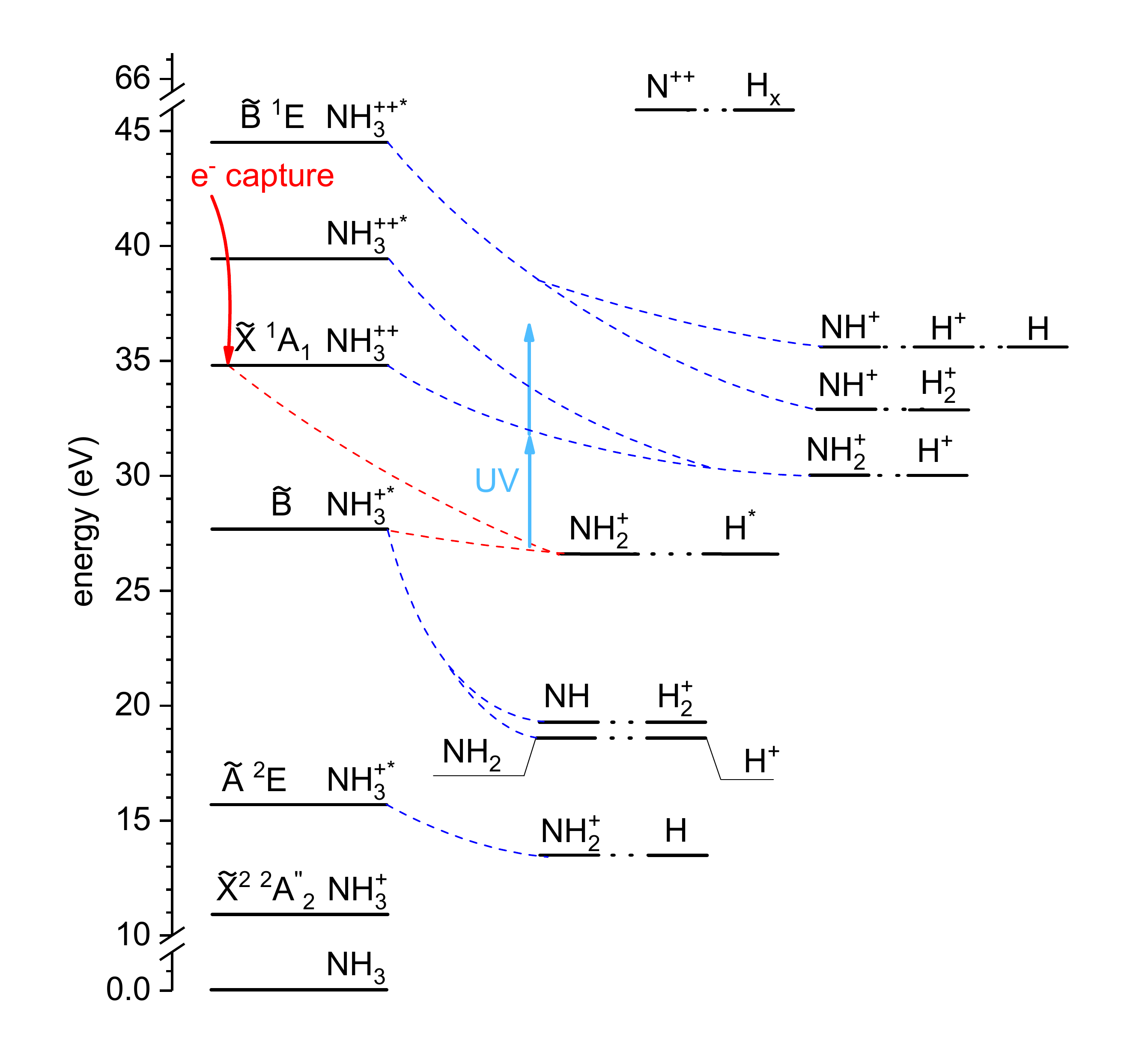}
		\caption{Schematic of the known excited states of NH$_3^+$ and NH$_3^{++}$, and their fragmentation channels (blue dotted lines) from~\cite{winkoun1986fragmentation,stankiewicz1989double,leyh1995reaction,langford1992determination}. For all shown states, the y-axis position reflects their thermodynamic energy threshold with respect to the NH$_3$ ground state. The red dotted line symbolizes the proposed new decay channel yielding excited hydrogen. Blue and red arrows show the electron capture from the doubly ionized state and the re-pumping with the UV probe laser. Please notice that the shown initial and final states are not complete and do not intend to show all possible state evolutions.}
		\label{fig:5}
	\end{center}
\end{figure}

With these presumptions, we next identify fragments that show an increase in yield during the third phase. This is the case for H$^{+}$ (Fig.\,\ref{fig:3}), N$^{++}$, NH$_2^+$ (Fig.\,\ref{fig:4}), and N$^{+}$ (Fig.\,S5\textsuperscript{\dag}). Of these three, only NH$_2^+$ has an overall negative probe yield, converging to zero for long pump-probe delays. As such, the NH$_2^+$ yield reproduces exactly the dynamic expected for a product produced by the undisturbed dissociation, i.e. without the probe laser. This observation restricts us to two possible parent molecules, NH$_3^{+*}$ and NH$_4^{+*}$, since we have H$^*$ + NH$_2^+$ as products. Complementary, we look for ions whose probe yield reduces in phase three. These ions are created by the UV laser lifting the excited molecule onto a different dissociative potential energy surface and/or ionizing it prior to dissociation. Only two ions show a decrease in probe ion yield in phase three. These two ions are H$_2^+$ and NH$^+$, shown in Fig.\,\ref{fig:5}. We notice that they match the products NH$_2^+$ and H$^*$ in numbers of H and N atoms. We thus propose the dissociation channel:
\begin{equation}NH_3^{+*}\rightarrow H^* + NH_2^+\end{equation} dissociating with a half-life time of $\tau$ = (1.6 $\pm$ 0.4)\,ps extracted from the energy shift of the photoelectrons. For pump-probe delays 700\,fs $<$ t $<$ 10\,ps, the UV laser interacting with the precursor molecule leads to different ions, partially governed by the reaction: \begin{equation}
NH_3^{+*} \xrightarrow{UV} NH^{+} + H_2^+ +e^-
\end{equation} 
This interpretation is consistent with the observed ion yields. Nonetheless, it is tentative, since in the observed system it is hardly possible to exclude all alternatives. Please note, that all of the above discussion is also valid if additional hydrogens are ejected as neutral fragments. In our experiment only charged fragments are detected, thus a differentiation of the possible parents NH$_3^{+*}$ from NH$_4^{+*}$ is not possible. Furthermore, the absolute changes in ion yield clearly show that Eq.\,2 is not the only possible fragmentation channel when ionizing NH$_3^{+*}$ prior to dissociation. This is not surprising, since a molecule with so much excess energy is very likely to have multiple dissociation pathways. A similar channel with NH$_2^{+*}$ as parent molecule, i.e. NH$_2^{+*} \rightarrow H^* + NH^+$ is possible and cannot be distinguished through the ion yields.\\
The dissociation of ammonia cations has been extensively studied using photoionization, ab-initio calculations,~\cite{winkoun1986fragmentation,samson1987absorption,viel2006photoionization,viel2008photoionization} as well as single-electron capture employing PEPIPICO.~\cite{leyh1995reaction} From these studies, the energetic dissociation threshold for NH$_3^{+*} \rightarrow$ NH$_2^+$+ H is known to be 15.76\,eV with respect to the ground state of ammonia, and is associated with the \~{A}$^2$E excited state of NH$_3^{+*}$. In a very simple approximation, adding 10.2\,eV excitation energy to reach H$^*$(n\,=\,2), one determines a minimum energy of 26\,eV required for the dissociation NH$_3^{+*} \rightarrow$ NH$_2^+$+ H$^*$(n\,=\,2). The double ionization threshold of ammonia is 33.7\,eV~\cite{stankiewicz1989double} and the \~{B} excited state of the ammonia cation can be reached when ionizing the 2a$_1$ valence shell of ammonia with 27.7\,eV.~\cite{banna1987study} As such, the dissociation yielding H$^*$ is energetically possible after electron capture by NH$_3^{++}$ and from the \~{B} excited state. Nonetheless, the possibility of a fragmentation channel producing excited hydrogen is not discussed in the literature, even though it is generally known that excited states of the amino radical fragment are produced.~\cite{florescu2006dissociative} In Fig.\,\ref{fig:5}, the most important states are shown in a highly simplified state scheme with the y-axis reflecting on the thermodynamic energy threshold for certain products. The proposed dissociation channel of NH$_3^{+*}$ is marked as dotted red lines, with dotted blue lines representing the known channels. We can see that after electron-capture from NH$_3^{++}$, dissociation into NH$_2^+$ and H$^*$ is energetically possible. Absorbing one or two photons from the UV probe laser, the dissociative energy surfaces coming from the doubly ionized ammonia can be reached. To explain our data, it is crucial that some of these channels do not create H$^+$ in their dissociation, as is the case when dissociating into NH$^+$ + H$_2^+$. The state scheme in Fig.\,\ref{fig:5} shows only the most important contributors and is neglecting, for example, doubly excited states, and many of the known dissociative recombination channels.~\cite{florescu2006dissociative} From the state scheme, our interpretation of excited hydrogen forming via dissociative recombination seems justified. Nonetheless, the overall high intensity in the channel is surprising. We assume that many of the dissociation channels governing the ion dynamics could not be identified in this work. It is known that dissociative recombination of nitrogen containing molecules often involves excited and ground-state neutrals~\cite{florescu2006dissociative} and it can be safely assumed that this is also possible for the very high lying states of NH$_3^{++*}$.\\

\section{Conclusions}
In this experiment, we have used an intense, XUV pulse to create a nanoplasma inside ammonia clusters. Using a time-delayed UV pulse, we probed the resulting excited states of molecules and atoms, thereby gaining information on their temporal evolution. We used simultaneous ion time-of-flight and electron velocity-map-imaging detection to map the formation of excited states and detect all charged fragments. Surprisingly, despite many possible secondary processes, the dominant fragment channels follow universal principles and were identified. The dynamics are governed by sequential one-photon ionization, electron-ion recombination, and electron-impact excitation creating various charged and neutral fragments. We characterize the cluster evolution after XUV irradiation as a three phase process. The first phase is dominated by rising cluster potential with fast electrons being trapped inside the ionic core. This is followed by the second phase, during which fast protons leave the cluster resulting in a drop of the electronic potential and detection of the UV-ionized electrons. In the third and final phase, the UV laser probes the molecular dissociation dynamics of excited and neutral cluster fragments. We observe two different formation channels for excited hydrogen. The first process is assigned to recombination of electrons and protons in the nanoplasma, whilst the second process is interpreted as dissociative recombination of doubly charged ammonia molecules with electrons. \\

\section*{Conflicts of interest}
``There are no conflicts to declare''.

\section*{Acknowledgements}
Funding from the Deutsche Forschungsgemeinschaft (STI 125/19-1, GRK 2079) Carl-Zeiss-Stiftung, grants 200021\_146598 and 200020\_162434 from the Swiss National Science Foundation, as well as the Department of Excellence, Department of Chemistry and Technologies of drugs, and Progetto Ateneo-2016 (prot. n. RG116154C8E02882) of Sapienza University are gratefully acknowledged.\\


\renewcommand\refname{References}
\bibliographystyle{rsc} 

\begin{mcitethebibliography}{45}
	\providecommand*{\natexlab}[1]{#1}
	\providecommand*{\mciteSetBstSublistMode}[1]{}
	\providecommand*{\mciteSetBstMaxWidthForm}[2]{}
	\providecommand*{\mciteBstWouldAddEndPuncttrue}
	{\def\EndOfBibitem{\unskip.}}
	\providecommand*{\mciteBstWouldAddEndPunctfalse}
	{\let\EndOfBibitem\relax}
	\providecommand*{\mciteSetBstMidEndSepPunct}[3]{}
	\providecommand*{\mciteSetBstSublistLabelBeginEnd}[3]{}
	\providecommand*{\EndOfBibitem}{}
	\mciteSetBstSublistMode{f}
	\mciteSetBstMaxWidthForm{subitem}
	{(\emph{\alph{mcitesubitemcount}})}
	\mciteSetBstSublistLabelBeginEnd{\mcitemaxwidthsubitemform\space}
	{\relax}{\relax}
	
	\bibitem[Fennel \emph{et~al.}(2010)Fennel, Meiwes-Broer, Tiggesb{\"a}umker,
	Reinhard, Dinh, and Suraud]{fennel2010laser}
	T.~Fennel, K.-H. Meiwes-Broer, J.~Tiggesb{\"a}umker, P.-G. Reinhard, P.~M. Dinh
	and E.~Suraud, \emph{Reviews of modern physics}, 2010, \textbf{82},
	1793\relax
	\mciteBstWouldAddEndPuncttrue
	\mciteSetBstMidEndSepPunct{\mcitedefaultmidpunct}
	{\mcitedefaultendpunct}{\mcitedefaultseppunct}\relax
	\EndOfBibitem
	\bibitem[Saalmann \emph{et~al.}(2006)Saalmann, Siedschlag, and
	Rost]{saalmann2006mechanisms}
	U.~Saalmann, C.~Siedschlag and J.~Rost, \emph{Journal of Physics B: Atomic,
		Molecular and Optical Physics}, 2006, \textbf{39}, R39\relax
	\mciteBstWouldAddEndPuncttrue
	\mciteSetBstMidEndSepPunct{\mcitedefaultmidpunct}
	{\mcitedefaultendpunct}{\mcitedefaultseppunct}\relax
	\EndOfBibitem
	\bibitem[Wabnitz \emph{et~al.}(2002)Wabnitz, Bittner, De~Castro, D{\"o}hrmann,
	G{\"u}rtler, Laarmann, Laasch, Schulz, Swiderski, von
	Haeften,\emph{et~al.}]{wabnitz2002multiple}
	H.~Wabnitz, L.~Bittner, A.~De~Castro, R.~D{\"o}hrmann, P.~G{\"u}rtler,
	T.~Laarmann, W.~Laasch, J.~Schulz, A.~Swiderski, K.~von Haeften
	\emph{et~al.}, \emph{Nature}, 2002, \textbf{420}, 482--485\relax
	\mciteBstWouldAddEndPuncttrue
	\mciteSetBstMidEndSepPunct{\mcitedefaultmidpunct}
	{\mcitedefaultendpunct}{\mcitedefaultseppunct}\relax
	\EndOfBibitem
	\bibitem[Bostedt \emph{et~al.}(2008)Bostedt, Thomas, Hoener, Eremina, Fennel,
	Meiwes-Broer, Wabnitz, Kuhlmann, Pl{\"o}njes,
	Tiedtke,\emph{et~al.}]{bostedt2008multistep}
	C.~Bostedt, H.~Thomas, M.~Hoener, E.~Eremina, T.~Fennel, K.-H. Meiwes-Broer,
	H.~Wabnitz, M.~Kuhlmann, E.~Pl{\"o}njes, K.~Tiedtke \emph{et~al.},
	\emph{Physical review letters}, 2008, \textbf{100}, 133401\relax
	\mciteBstWouldAddEndPuncttrue
	\mciteSetBstMidEndSepPunct{\mcitedefaultmidpunct}
	{\mcitedefaultendpunct}{\mcitedefaultseppunct}\relax
	\EndOfBibitem
	\bibitem[Arbeiter and Fennel(2011)]{arbeiter2011rare}
	M.~Arbeiter and T.~Fennel, \emph{New Journal of Physics}, 2011, \textbf{13},
	053022\relax
	\mciteBstWouldAddEndPuncttrue
	\mciteSetBstMidEndSepPunct{\mcitedefaultmidpunct}
	{\mcitedefaultendpunct}{\mcitedefaultseppunct}\relax
	\EndOfBibitem
	\bibitem[Arbeiter and Fennel(2010)]{arbeiter2010ionization}
	M.~Arbeiter and T.~Fennel, \emph{Physical Review A}, 2010, \textbf{82},
	013201\relax
	\mciteBstWouldAddEndPuncttrue
	\mciteSetBstMidEndSepPunct{\mcitedefaultmidpunct}
	{\mcitedefaultendpunct}{\mcitedefaultseppunct}\relax
	\EndOfBibitem
	\bibitem[Iwayama \emph{et~al.}(2013)Iwayama, Nagaya, Yao, Fukuzawa, Liu,
	Pr{\"u}mper, Motomura, Ueda, Saito,
	Rudenko,\emph{et~al.}]{iwayama2013frustration}
	H.~Iwayama, K.~Nagaya, M.~Yao, H.~Fukuzawa, X.~Liu, G.~Pr{\"u}mper,
	K.~Motomura, K.~Ueda, N.~Saito, A.~Rudenko \emph{et~al.}, \emph{Journal of
		Physics B: Atomic, Molecular and Optical Physics}, 2013, \textbf{46},
	164019\relax
	\mciteBstWouldAddEndPuncttrue
	\mciteSetBstMidEndSepPunct{\mcitedefaultmidpunct}
	{\mcitedefaultendpunct}{\mcitedefaultseppunct}\relax
	\EndOfBibitem
	\bibitem[Jungreuthmayer \emph{et~al.}(2005)Jungreuthmayer, Ramunno,
	Zanghellini, and Brabec]{jungreuthmayer2005intense}
	C.~Jungreuthmayer, L.~Ramunno, J.~Zanghellini and T.~Brabec, \emph{Journal of
		Physics B: Atomic, Molecular and Optical Physics}, 2005, \textbf{38},
	3029\relax
	\mciteBstWouldAddEndPuncttrue
	\mciteSetBstMidEndSepPunct{\mcitedefaultmidpunct}
	{\mcitedefaultendpunct}{\mcitedefaultseppunct}\relax
	\EndOfBibitem
	\bibitem[M{\"u}ller \emph{et~al.}(2015)M{\"u}ller, Schroedter, Oelze,
	N{\"o}sel, Przystawik, Kickermann, Adolph, Gorkhover, Fl{\"u}ckiger,
	Krikunova,\emph{et~al.}]{muller2015ionization}
	M.~M{\"u}ller, L.~Schroedter, T.~Oelze, L.~N{\"o}sel, A.~Przystawik,
	A.~Kickermann, M.~Adolph, T.~Gorkhover, L.~Fl{\"u}ckiger, M.~Krikunova
	\emph{et~al.}, \emph{Journal of Physics B: Atomic, Molecular and Optical
		Physics}, 2015, \textbf{48}, 174002\relax
	\mciteBstWouldAddEndPuncttrue
	\mciteSetBstMidEndSepPunct{\mcitedefaultmidpunct}
	{\mcitedefaultendpunct}{\mcitedefaultseppunct}\relax
	\EndOfBibitem
	\bibitem[Sch{\"u}tte \emph{et~al.}(2014)Sch{\"u}tte, Campi, Arbeiter, Fennel,
	Vrakking, and Rouz{\'e}e]{schutte2014tracing}
	B.~Sch{\"u}tte, F.~Campi, M.~Arbeiter, T.~Fennel, M.~Vrakking and
	A.~Rouz{\'e}e, \emph{Physical review letters}, 2014, \textbf{112},
	253401\relax
	\mciteBstWouldAddEndPuncttrue
	\mciteSetBstMidEndSepPunct{\mcitedefaultmidpunct}
	{\mcitedefaultendpunct}{\mcitedefaultseppunct}\relax
	\EndOfBibitem
	\bibitem[Oelze \emph{et~al.}(2017)Oelze, Sch{\"u}tte, M{\"u}ller, M{\"u}ller,
	Wieland, Fr{\"u}hling, Drescher, Al-Shemmary, Golz,
	Stojanovic,\emph{et~al.}]{oelze2017correlated}
	T.~Oelze, B.~Sch{\"u}tte, M.~M{\"u}ller, J.~P. M{\"u}ller, M.~Wieland,
	U.~Fr{\"u}hling, M.~Drescher, A.~Al-Shemmary, T.~Golz, N.~Stojanovic
	\emph{et~al.}, \emph{Scientific reports}, 2017, \textbf{7}, 40736\relax
	\mciteBstWouldAddEndPuncttrue
	\mciteSetBstMidEndSepPunct{\mcitedefaultmidpunct}
	{\mcitedefaultendpunct}{\mcitedefaultseppunct}\relax
	\EndOfBibitem
	\bibitem[Sch{\"u}tte \emph{et~al.}(2017)Sch{\"u}tte, Vrakking, and
	Rouz{\'e}e]{schutte2017tracing}
	B.~Sch{\"u}tte, M.~J. Vrakking and A.~Rouz{\'e}e, \emph{Physical Review A},
	2017, \textbf{95}, 063417\relax
	\mciteBstWouldAddEndPuncttrue
	\mciteSetBstMidEndSepPunct{\mcitedefaultmidpunct}
	{\mcitedefaultendpunct}{\mcitedefaultseppunct}\relax
	\EndOfBibitem
	\bibitem[Pohl \emph{et~al.}(2008)Pohl, Vrinceanu, and
	Sadeghpour]{pohl2008rydberg}
	T.~Pohl, D.~Vrinceanu and H.~Sadeghpour, \emph{Physical review letters}, 2008,
	\textbf{100}, 223201\relax
	\mciteBstWouldAddEndPuncttrue
	\mciteSetBstMidEndSepPunct{\mcitedefaultmidpunct}
	{\mcitedefaultendpunct}{\mcitedefaultseppunct}\relax
	\EndOfBibitem
	\bibitem[Iwayama \emph{et~al.}(2015)Iwayama, Harries, and
	Shigemasa]{iwayama2015transient}
	H.~Iwayama, J.~Harries and E.~Shigemasa, \emph{Physical Review A}, 2015,
	\textbf{91}, 021402\relax
	\mciteBstWouldAddEndPuncttrue
	\mciteSetBstMidEndSepPunct{\mcitedefaultmidpunct}
	{\mcitedefaultendpunct}{\mcitedefaultseppunct}\relax
	\EndOfBibitem
	\bibitem[Sch{\"u}tte \emph{et~al.}(2015)Sch{\"u}tte, Oelze, Krikunova,
	Arbeiter, Fennel, Vrakking, and Rouz{\'e}e]{schutte2015real}
	B.~Sch{\"u}tte, T.~Oelze, M.~Krikunova, M.~Arbeiter, T.~Fennel, M.~J. Vrakking
	and A.~Rouz{\'e}e, \emph{Journal of Physics B: Atomic, Molecular and Optical
		Physics}, 2015, \textbf{48}, 185101\relax
	\mciteBstWouldAddEndPuncttrue
	\mciteSetBstMidEndSepPunct{\mcitedefaultmidpunct}
	{\mcitedefaultendpunct}{\mcitedefaultseppunct}\relax
	\EndOfBibitem
	\bibitem[Di~Cintio \emph{et~al.}(2013)Di~Cintio, Saalmann, and
	Rost]{di2013proton}
	P.~Di~Cintio, U.~Saalmann and J.-M. Rost, \emph{Physical review letters}, 2013,
	\textbf{111}, 123401\relax
	\mciteBstWouldAddEndPuncttrue
	\mciteSetBstMidEndSepPunct{\mcitedefaultmidpunct}
	{\mcitedefaultendpunct}{\mcitedefaultseppunct}\relax
	\EndOfBibitem
	\bibitem[Kaplan \emph{et~al.}(2003)Kaplan, Dubetsky, and
	Shkolnikov]{kaplan2003shock}
	A.~E. Kaplan, B.~Y. Dubetsky and P.~Shkolnikov, \emph{Physical review letters},
	2003, \textbf{91}, 143401\relax
	\mciteBstWouldAddEndPuncttrue
	\mciteSetBstMidEndSepPunct{\mcitedefaultmidpunct}
	{\mcitedefaultendpunct}{\mcitedefaultseppunct}\relax
	\EndOfBibitem
	\bibitem[Last and Jortner(2001)]{last2001nuclear}
	I.~Last and J.~Jortner, \emph{Physical Review A}, 2001, \textbf{64},
	063201\relax
	\mciteBstWouldAddEndPuncttrue
	\mciteSetBstMidEndSepPunct{\mcitedefaultmidpunct}
	{\mcitedefaultendpunct}{\mcitedefaultseppunct}\relax
	\EndOfBibitem
	\bibitem[Snyder \emph{et~al.}(1996)Snyder, Wei, Purnell, Buzza, and
	Castleman~Jr]{snyder1996femtosecond}
	E.~Snyder, S.~Wei, J.~Purnell, S.~Buzza and A.~Castleman~Jr, \emph{Chemical
		physics letters}, 1996, \textbf{248}, 1--7\relax
	\mciteBstWouldAddEndPuncttrue
	\mciteSetBstMidEndSepPunct{\mcitedefaultmidpunct}
	{\mcitedefaultendpunct}{\mcitedefaultseppunct}\relax
	\EndOfBibitem
	\bibitem[Card \emph{et~al.}(1997)Card, Folmer, Sato, Buzza, and
	Castleman]{card1997covariance}
	D.~Card, D.~Folmer, S.~Sato, S.~Buzza and A.~Castleman, \emph{The Journal of
		Physical Chemistry A}, 1997, \textbf{101}, 3417--3423\relax
	\mciteBstWouldAddEndPuncttrue
	\mciteSetBstMidEndSepPunct{\mcitedefaultmidpunct}
	{\mcitedefaultendpunct}{\mcitedefaultseppunct}\relax
	\EndOfBibitem
	\bibitem[Niu \emph{et~al.}(2005)Niu, Li, Liang, Wen, and
	Luo]{niu2005covariance}
	D.~Niu, H.~Li, F.~Liang, L.~Wen and X.~Luo, \emph{Chinese science bulletin},
	2005, \textbf{50}, 2115--2117\relax
	\mciteBstWouldAddEndPuncttrue
	\mciteSetBstMidEndSepPunct{\mcitedefaultmidpunct}
	{\mcitedefaultendpunct}{\mcitedefaultseppunct}\relax
	\EndOfBibitem
	\bibitem[Wang \emph{et~al.}(2008)Wang, Li, Niu, Wen, and
	Zhang]{wang2008cluster}
	W.~Wang, H.~Li, D.~Niu, L.~Wen and N.~Zhang, \emph{Chemical Physics}, 2008,
	\textbf{352}, 111--116\relax
	\mciteBstWouldAddEndPuncttrue
	\mciteSetBstMidEndSepPunct{\mcitedefaultmidpunct}
	{\mcitedefaultendpunct}{\mcitedefaultseppunct}\relax
	\EndOfBibitem
	\bibitem[Iwan \emph{et~al.}(2012)Iwan, Andreasson, Bergh, Schorb, Thomas, Rupp,
	Gorkhover, Adolph, M{\"o}ller, Bostedt,\emph{et~al.}]{iwan2012explosion}
	B.~Iwan, J.~Andreasson, M.~Bergh, S.~Schorb, H.~Thomas, D.~Rupp, T.~Gorkhover,
	M.~Adolph, T.~M{\"o}ller, C.~Bostedt \emph{et~al.}, \emph{Physical Review A},
	2012, \textbf{86}, 033201\relax
	\mciteBstWouldAddEndPuncttrue
	\mciteSetBstMidEndSepPunct{\mcitedefaultmidpunct}
	{\mcitedefaultendpunct}{\mcitedefaultseppunct}\relax
	\EndOfBibitem
	\bibitem[Lyamayev \emph{et~al.}(2013)Lyamayev, Ovcharenko, Katzy, Devetta,
	Bruder, LaForge, Mudrich, Person, Stienkemeier,
	Krikunova,\emph{et~al.}]{lyamayev2013modular}
	V.~Lyamayev, Y.~Ovcharenko, R.~Katzy, M.~Devetta, L.~Bruder, A.~LaForge,
	M.~Mudrich, U.~Person, F.~Stienkemeier, M.~Krikunova \emph{et~al.},
	\emph{Journal of Physics B: Atomic, Molecular and Optical Physics}, 2013,
	\textbf{46}, 164007\relax
	\mciteBstWouldAddEndPuncttrue
	\mciteSetBstMidEndSepPunct{\mcitedefaultmidpunct}
	{\mcitedefaultendpunct}{\mcitedefaultseppunct}\relax
	\EndOfBibitem
	\bibitem[Allaria \emph{et~al.}(2012)Allaria, Appio, Badano, Barletta,
	Bassanese, Biedron, Borga, Busetto, Castronovo,
	Cinquegrana,\emph{et~al.}]{allaria2012highly}
	E.~Allaria, R.~Appio, L.~Badano, W.~Barletta, S.~Bassanese, S.~Biedron,
	A.~Borga, E.~Busetto, D.~Castronovo, P.~Cinquegrana \emph{et~al.},
	\emph{Nature Photonics}, 2012, \textbf{6}, 699\relax
	\mciteBstWouldAddEndPuncttrue
	\mciteSetBstMidEndSepPunct{\mcitedefaultmidpunct}
	{\mcitedefaultendpunct}{\mcitedefaultseppunct}\relax
	\EndOfBibitem
	\bibitem[Allaria \emph{et~al.}(2012)Allaria, Battistoni, Bencivenga, Borghes,
	Callegari, Capotondi, Castronovo, Cinquegrana, Cocco,
	Coreno,\emph{et~al.}]{allaria2012tunability}
	E.~Allaria, A.~Battistoni, F.~Bencivenga, R.~Borghes, C.~Callegari,
	F.~Capotondi, D.~Castronovo, P.~Cinquegrana, D.~Cocco, M.~Coreno
	\emph{et~al.}, \emph{New Journal of Physics}, 2012, \textbf{14}, 113009\relax
	\mciteBstWouldAddEndPuncttrue
	\mciteSetBstMidEndSepPunct{\mcitedefaultmidpunct}
	{\mcitedefaultendpunct}{\mcitedefaultseppunct}\relax
	\EndOfBibitem
	\bibitem[Bobbert \emph{et~al.}(2002)Bobbert, Sch{\"u}tte, Steinbach, and
	Buck]{bobbert2002fragmentation}
	C.~Bobbert, S.~Sch{\"u}tte, C.~Steinbach and U.~Buck, \emph{The European
		Physical Journal D-Atomic, Molecular, Optical and Plasma Physics}, 2002,
	\textbf{19}, 183--192\relax
	\mciteBstWouldAddEndPuncttrue
	\mciteSetBstMidEndSepPunct{\mcitedefaultmidpunct}
	{\mcitedefaultendpunct}{\mcitedefaultseppunct}\relax
	\EndOfBibitem
	\bibitem[Dick(2014)]{dick2014inverting}
	B.~Dick, \emph{Physical Chemistry Chemical Physics}, 2014, \textbf{16},
	570--580\relax
	\mciteBstWouldAddEndPuncttrue
	\mciteSetBstMidEndSepPunct{\mcitedefaultmidpunct}
	{\mcitedefaultendpunct}{\mcitedefaultseppunct}\relax
	\EndOfBibitem
	\bibitem[Samson \emph{et~al.}(1987)Samson, Haddad, and
	Kilcoyne]{samson1987absorption}
	J.~A. Samson, G.~Haddad and L.~Kilcoyne, \emph{The Journal of chemical
		physics}, 1987, \textbf{87}, 6416--6422\relax
	\mciteBstWouldAddEndPuncttrue
	\mciteSetBstMidEndSepPunct{\mcitedefaultmidpunct}
	{\mcitedefaultendpunct}{\mcitedefaultseppunct}\relax
	\EndOfBibitem
	\bibitem[Li and Varandas(2010)]{li2010ab}
	Y.~Li and A.~Varandas, \emph{The Journal of Physical Chemistry A}, 2010,
	\textbf{114}, 6669--6680\relax
	\mciteBstWouldAddEndPuncttrue
	\mciteSetBstMidEndSepPunct{\mcitedefaultmidpunct}
	{\mcitedefaultendpunct}{\mcitedefaultseppunct}\relax
	\EndOfBibitem
	\bibitem[Biesner \emph{et~al.}(1989)Biesner, Schnieder, Ahlers, Xie, Welge,
	Ashfold, and Dixon]{biesner1989state}
	J.~Biesner, L.~Schnieder, G.~Ahlers, X.~Xie, K.~Welge, M.~Ashfold and R.~Dixon,
	\emph{The Journal of chemical physics}, 1989, \textbf{91}, 2901--2911\relax
	\mciteBstWouldAddEndPuncttrue
	\mciteSetBstMidEndSepPunct{\mcitedefaultmidpunct}
	{\mcitedefaultendpunct}{\mcitedefaultseppunct}\relax
	\EndOfBibitem
	\bibitem[Bach \emph{et~al.}(2003)Bach, Hutchison, Holiday, and
	Crim]{bach2003competition}
	A.~Bach, J.~M. Hutchison, R.~J. Holiday and F.~F. Crim, \emph{The Journal of
		Physical Chemistry A}, 2003, \textbf{107}, 10490--10496\relax
	\mciteBstWouldAddEndPuncttrue
	\mciteSetBstMidEndSepPunct{\mcitedefaultmidpunct}
	{\mcitedefaultendpunct}{\mcitedefaultseppunct}\relax
	\EndOfBibitem
	\bibitem[Freudenberg \emph{et~al.}(1996)Freudenberg, Radloff, Ritze, Stert,
	Weyers, Noack, and Hertel]{freudenberg1996ultrafast}
	T.~Freudenberg, W.~Radloff, H.-H. Ritze, V.~Stert, K.~Weyers, F.~Noack and
	I.~Hertel, \emph{Zeitschrift f{\"u}r Physik D Atoms, Molecules and Clusters},
	1996, \textbf{36}, 349--364\relax
	\mciteBstWouldAddEndPuncttrue
	\mciteSetBstMidEndSepPunct{\mcitedefaultmidpunct}
	{\mcitedefaultendpunct}{\mcitedefaultseppunct}\relax
	\EndOfBibitem
	\bibitem[LaForge \emph{et~al.}(2019)LaForge, Michiels, Bohlen, Callegari,
	Clark, von Conta, Coreno, Di~Fraia, Drabbels,
	Huppert,\emph{et~al.}]{laforge2019real}
	A.~C. LaForge, R.~Michiels, M.~Bohlen, C.~Callegari, A.~Clark, A.~von Conta,
	M.~Coreno, M.~Di~Fraia, M.~Drabbels, M.~Huppert \emph{et~al.}, \emph{Physical
		review letters}, 2019, \textbf{122}, 133001\relax
	\mciteBstWouldAddEndPuncttrue
	\mciteSetBstMidEndSepPunct{\mcitedefaultmidpunct}
	{\mcitedefaultendpunct}{\mcitedefaultseppunct}\relax
	\EndOfBibitem
	\bibitem[Bates(2012)]{bates2012atomic}
	D.~R. Bates, \emph{Atomic and molecular processes}, Elsevier, 2012,
	vol.~13\relax
	\mciteBstWouldAddEndPuncttrue
	\mciteSetBstMidEndSepPunct{\mcitedefaultmidpunct}
	{\mcitedefaultendpunct}{\mcitedefaultseppunct}\relax
	\EndOfBibitem
	\bibitem[Burgess(1965)]{burgess1965tables}
	A.~Burgess, \emph{Memoirs of the Royal Astronomical Society}, 1965,
	\textbf{69}, 1\relax
	\mciteBstWouldAddEndPuncttrue
	\mciteSetBstMidEndSepPunct{\mcitedefaultmidpunct}
	{\mcitedefaultendpunct}{\mcitedefaultseppunct}\relax
	\EndOfBibitem	
	\bibitem[Davies \emph{et~al.}(1999)Davies, LeClaire, Continetti, and
	Hayden]{davies1999femtosecond}
	J.~Davies, J.~LeClaire, R.~Continetti and C.~Hayden, \emph{The Journal of
		chemical physics}, 1999, \textbf{111}, 1--4\relax
	\mciteBstWouldAddEndPuncttrue
	\mciteSetBstMidEndSepPunct{\mcitedefaultmidpunct}
	{\mcitedefaultendpunct}{\mcitedefaultseppunct}\relax
	\EndOfBibitem
	\bibitem[Meier and Engel(2002)]{meier2002time}
	C.~Meier and V.~Engel, \emph{Physical Chemistry Chemical Physics}, 2002,
	\textbf{4}, 5014--5019\relax
	\mciteBstWouldAddEndPuncttrue
	\mciteSetBstMidEndSepPunct{\mcitedefaultmidpunct}
	{\mcitedefaultendpunct}{\mcitedefaultseppunct}\relax
	\EndOfBibitem
	\bibitem[Tsubouchi \emph{et~al.}(2005)Tsubouchi, de~Lange, and
	Suzuki]{tsubouchi2005ultrafast}
	M.~Tsubouchi, C.~A. de~Lange and T.~Suzuki, \emph{Journal of electron
		spectroscopy and related phenomena}, 2005, \textbf{142}, 193--205\relax
	\mciteBstWouldAddEndPuncttrue
	\mciteSetBstMidEndSepPunct{\mcitedefaultmidpunct}
	{\mcitedefaultendpunct}{\mcitedefaultseppunct}\relax
	\EndOfBibitem
	\bibitem[Winkoun and Dujardin(1986)]{winkoun1986fragmentation}
	D.~Winkoun and G.~Dujardin, \emph{Zeitschrift f{\"u}r Physik D Atoms, Molecules
		and Clusters}, 1986, \textbf{4}, 57--64\relax
	\mciteBstWouldAddEndPuncttrue
	\mciteSetBstMidEndSepPunct{\mcitedefaultmidpunct}
	{\mcitedefaultendpunct}{\mcitedefaultseppunct}\relax
	\EndOfBibitem
	\bibitem[Viel \emph{et~al.}(2006)Viel, Eisfeld, Neumann, Domcke, and
	Manthe]{viel2006photoionization}
	A.~Viel, W.~Eisfeld, S.~Neumann, W.~Domcke and U.~Manthe, \emph{The Journal of
		chemical physics}, 2006, \textbf{124}, 214306\relax
	\mciteBstWouldAddEndPuncttrue
	\mciteSetBstMidEndSepPunct{\mcitedefaultmidpunct}
	{\mcitedefaultendpunct}{\mcitedefaultseppunct}\relax
	\EndOfBibitem
	\bibitem[Viel \emph{et~al.}(2008)Viel, Eisfeld, Evenhuis, and
	Manthe]{viel2008photoionization}
	A.~Viel, W.~Eisfeld, C.~R. Evenhuis and U.~Manthe, \emph{Chemical Physics},
	2008, \textbf{347}, 331--339\relax
	\mciteBstWouldAddEndPuncttrue
	\mciteSetBstMidEndSepPunct{\mcitedefaultmidpunct}
	{\mcitedefaultendpunct}{\mcitedefaultseppunct}\relax
	\EndOfBibitem
	\bibitem[Leyh and Hoxha(1995)]{leyh1995reaction}
	B.~Leyh and A.~Hoxha, \emph{Chemical physics}, 1995, \textbf{192}, 65--77\relax
	\mciteBstWouldAddEndPuncttrue
	\mciteSetBstMidEndSepPunct{\mcitedefaultmidpunct}
	{\mcitedefaultendpunct}{\mcitedefaultseppunct}\relax
	\EndOfBibitem
	\bibitem[Stankiewicz \emph{et~al.}(1989)Stankiewicz, Hatherly, Frasinski,
	Codling, and Holland]{stankiewicz1989double}
	M.~Stankiewicz, P.~Hatherly, L.~Frasinski, K.~Codling and D.~Holland,
	\emph{Journal of Physics B: Atomic, Molecular and Optical Physics}, 1989,
	\textbf{22}, 21\relax
	\mciteBstWouldAddEndPuncttrue
	\mciteSetBstMidEndSepPunct{\mcitedefaultmidpunct}
	{\mcitedefaultendpunct}{\mcitedefaultseppunct}\relax
	\EndOfBibitem
	\bibitem[Banna \emph{et~al.}(1987)Banna, Kossmann, and Schmidt]{banna1987study}
	M.~S. Banna, H.~Kossmann and V.~Schmidt, \emph{Chemical physics}, 1987,
	\textbf{114}, 157--163\relax
	\mciteBstWouldAddEndPuncttrue
	\mciteSetBstMidEndSepPunct{\mcitedefaultmidpunct}
	{\mcitedefaultendpunct}{\mcitedefaultseppunct}\relax
	\EndOfBibitem
	\bibitem[Florescu-Mitchell and Mitchell(2006)]{florescu2006dissociative}
	A.~Florescu-Mitchell and J.~Mitchell, \emph{Physics reports}, 2006,
	\textbf{430}, 277--374\relax
	\mciteBstWouldAddEndPuncttrue
	\mciteSetBstMidEndSepPunct{\mcitedefaultmidpunct}
	{\mcitedefaultendpunct}{\mcitedefaultseppunct}\relax
	\EndOfBibitem
	\bibitem[Langford \emph{et~al.}(1992)Langford, Harris, Fournier, and
	Fournier]{langford1992determination}
	M.~Langford, F.~Harris, P.~Fournier and J.~Fournier, \emph{International
		journal of mass spectrometry and ion processes}, 1992, \textbf{116},
	53--69\relax
	\mciteBstWouldAddEndPuncttrue
	\mciteSetBstMidEndSepPunct{\mcitedefaultmidpunct}
	{\mcitedefaultendpunct}{\mcitedefaultseppunct}\relax
	\EndOfBibitem
\end{mcitethebibliography}
\providecommand*{\mcitethebibliography}{\thebibliography}
\csname @ifundefined\endcsname{endmcitethebibliography}
{\let\endmcitethebibliography\endthebibliography}{}

\end{document}